\documentclass[aps,prl,12pt]{revtex4}
\usepackage{epsf}

\begin{document}

\epsfverbosetrue
\begin{titlepage}
\begin{center}
{\LARGE {\bf{Bragg-grating solitons in a semilinear dual-core system}}
} \\     \vspace{12mm}
{\large \rm Javid Atai$^1$ and Boris A. Malomed$^2$}
\\ \vspace{3mm}
{\it
$^{(1)}$School of Electrical and Information Engineering,
The University of Sydney,
Sydney, NSW 2006, Australia
} \\ \vspace{1.5mm}
{\it
$^{(2)}$Department of Interdisciplinary Studies,
Faculty of Engineering, Tel Aviv University,
Tel Aviv 69978, Israel}
\vspace{1.5mm}
\end{center}
\vspace{18mm}
\section*{abstract}
We investigate the existence and stability of
gap solitons in a double-core optical fiber,
where one core has the Kerr
nonlinearity and the other one is linear, with
the Bragg grating (BG)
written on the nonlinear core, while the linear one
may or may not have BG. The model considerably extends
the previously studied families
of BG solitons. For zero-velocity solitons, we
find exact solutions in a limiting case when
the  group-velocity terms are absent in the equation for
the linear core. In the general case,
solitons are found numerically.
Stability borders for the
solitons are found in terms of an internal parameter of the
soliton family. Depending on frequency 
$\omega$, the solitons may remain stable for  
large values of the group velocity in 
 the linear core.  Stable moving
solitons are found too. They are produced by
interaction of initially separated solitons,
which shows a considerable spontaneous
symmetry breaking in the case when the solitons
attract each other.
\end{titlepage}


\section{Introduction and formulation of the model}

It is well known that the combination of the Kerr nonlinearity with a strong
effective dispersion induced by the resonant reflection of light on the
Bragg grating (BG) gives rise to a vast family of gap solitons, frequently
called BG solitons \cite{review} (in this work, we use the term ``soliton''
in the loose sense, without implying integrability of the model where it
appears; in particular, it will be shown that interactions between
``solitons'' in a model to be introduced below may be essentially
inelastic). A generally accepted mathematical model of the nonlinear fiber
equipped with BG is the so-called generalized massive Thirring model (GMTM) 
\cite{AW}. Thorough theoretical investigation of BG solitons, an important
step in which was the discovery of a class of exact single-soliton solutions
to GMTM \cite{AW}, was followed by observation of BG solitons created by a
very strong laser pulse launched into in a short segment ($\sim 6$ cm) of a
nonlinear optical fiber with the resonant BG written on it \cite{exper}.
Experimental studies of BG solitons were further developed (including, in
particular, formation of multiple BG solitons) in Refs. \cite{exper2}.

Observation of solitons in such a short fiber paves the way for many
potential applications, as well as for new experiments aimed at the study of
fundamental properties of optical solitons. This also makes it relevant to
consider more sophisticated nonlinear systems based on fiber gratings, where
the properties of solitons might still be more promising. In particular, one
can look for solitons in a {\it dual-core} system with linear coupling
between the cores, BG being written on both cores or a single one. The case
of two identical BG-carrying cores was considered in Ref. \cite{MakSymm},
where it was found that the model gave rise to a {\it bifurcation} at a
critical value of the soliton's energy. The bifurcation destabilizes a
symmetric two-component solution, simultaneously generating a nontrivial
asymmetric soliton. A dual-fiber system with unlike cores is easier to
fabricate and may offer new possibilities. One of the most interesting dual
systems with different cores is a {\it semilinear} one, where one core  is
linear. Semilinear dual-core models without BG were introduced earlier; both
continuous-wave and soliton states in them have been studied in various
contexts\cite{cw,MakAsymm}.

The objective of this work is to introduce a semilinear dual-core model in
which BG is written either on the nonlinear core only or on both cores, and
to search for solitons in it (which makes it necessary, first of all, to
explore the system's linear spectrum). Following the derivation of GMTM \cite
{review} and of the standard equations for a dual-core fiber (see, e.g.,
Ref. \cite{dual}) from Maxwell's equations, a general model for the
semilinear dual-core BG-equipped system can be written as the following set
of normalized equations: 
\begin{eqnarray}
iu_{t}+iu_{x}+\left[ |v|^{2}+(1/2)|u|^{2}\right] u+v+\kappa \phi &=&0,
\label{u} \\
iv_{t}-iv_{x}+\left[ |u|^{2}+(1/2)|v|^{2}\right] v+u+\kappa \psi &=&0,
\label{v}
\end{eqnarray}
\begin{eqnarray}
&&i\phi _{t}+ic\phi _{x}+\kappa u+\left( \lambda +i\mu \right) \psi =0,
\label{phi} \\
&&i\psi _{t}-ic\psi _{x}+\kappa v+\left( \lambda -i\mu \right) \phi =0
\label{psi}
\end{eqnarray}
Here, $u$ and $v$ represent the forward- and backward-propagating waves in
the nonlinear core, $\phi $ and $\psi $ are their counterparts in the linear
one, $\kappa $ is the coefficient of linear coupling between the cores,
while $\lambda $ and $\mu $ are the real and imaginary parts of the
BG-coupling coefficient in the linear core (which is, generally, complex if
its counterpart in the nonlinear core is normalized to be $1$, as is the
case in Eqs. (\ref{u}) and (\ref{v})). Lastly, the group velocity in the
nonlinear core is set equal to $1$, and $c${\bf \ }is the relative group
velocity in the linear core.

The simplest case giving rise to a novel system is $\lambda =\mu =0$
(corresponding to the linear core without BG), while cross-core coupling $%
\kappa $ is nonzero. Below, we will always set $\mu =0$; in most cases, $%
\lambda $ will also be zero, but effects of $\lambda \neq 0$ on the
solitons' stability will be investigated too. Note that, although the
present model finds its most natural formulation in the temporal domain, it
can also be readily interpreted in terms of the {\it spatial-domain}
evolution of the fields in a two-core planar waveguide, BG being realized as
a system of parallel cores written on the waveguide(s) \cite{spatial}.

It may also be quite interesting to consider a system where the Kerr
nonlinearity and BG are separated, i.e., with the grating written only on
the {\em linear} core. The corresponding model is obtained from the above
equations, dropping the linear terms $v$ and $u$ in Eqs. (\ref{u}) and (\ref
{v}) and setting $\lambda =1$ and $\mu =0$ in Eqs. (\ref{phi}) and (\ref{psi}%
). This model, which also seems quite promising, will be considered
elsewhere.

Before looking for solitons, it is necessary to analyze the spectrum of the
linearized system, in order to identify a spectral {\it gap} in which BG
solitons may reside \cite{review}. For a linear wave $\sim \exp \left(
ikx-i\omega t\right) $ and setting $\mu =0$, a dispersion equation for $%
\omega (k)$ can be obtained: 
\[
\omega ^{4}-\left[ 1+2\kappa ^{2}+\lambda ^{2}+\left( 1+c^{2}\right) k^{2}%
\right] \omega ^{2}+\left( \lambda -\kappa ^{2}\right) ^{2}+\left(
c^{2}-2c\kappa ^{2}+\lambda ^{2}\right) k^{2}+c^{2}k^{4}=0. 
\]
Analyzing this equation, it is easy to conclude that the gap does {\em not}
exist in the present model if $\lambda <\kappa ^{2}$ and $c^{2}-\lambda
+\lambda ^{2}<(2c-1)\kappa ^{2}$, or if $(1+2c)^{-1}\left( c+c^{2}+\lambda
^{2}\right) <\kappa ^{2}<\lambda $. In all the other cases, a finite gap is
present, and BG solitons may exist. In the particular case when the linear
properties of the two cores are identical, i.e., $c=1$ and $\lambda =1$,
which physically corresponds to identical BGs written on them, the gap
existence condition takes a very simple form, $\kappa ^{2}<1$ \cite{MakSymm}.

A remarkable property of the above-mentioned GMTM equations, to which Eqs.
(1) -~(4) reduce if the additional core is dropped, is the availability of
exact single-soliton solutions, both quiescent and moving with an arbitrary
velocity $v$, limited by $|v|<1$, despite the fact that the model is not
integrable (except for the unphysical case when the self-phase modulation
terms are omitted) \cite{AW}. Here, we aim to find soliton solutions to the
full system (1) - (4) and investigate their stability and interactions.
Solitons with zero velocity will be studied in detail, and moving solitons
will be presented too. In fact, the existence of the solitons with zero
velocity (which have not yet been observed experimentally in the single-core
fiber gratings) is the most intriguing possibility, as this implies a
possibility of ``full stoppage of light'' through its dynamical trapping,
which is especially interesting in view of the recent discovery of
``ultraslow light'' in ultracold gases \cite{slow}.

As for the physical parameters of the system and its soliton solutions, a
crucial factor is the ratio of the length $z_{{\rm coupl}}$ of the coupling
between the cores and a characteristic propagation distance (soliton's {\it %
dispersion length}) $z_{{\rm sol}}$ necessary for the formation of a soliton
in a single-core fiber with BG. As is well known, the former length in
available dual-core fibers is, normally, $\sim 1$ cm, and, according to the
experimental data \cite{exper,exper2}, $z_{{\rm sol}}$ is on the same order
of magnitude (it is so short, despite the fact that the solitons are
relatively broad in the temporal domain, because BG gives rise to an
extremely strong effective dispersion). This circumstance, $z_{{\rm coupl}%
}\sim z_{{\rm sol}}$, is quite favorable, as it suggests that the interplay
between the resonant light reflection on BG, Kerr nonlinearity, and linear
coupling between the cores may give rise to solitons with fairly novel
properties, in comparison with both the usual (single-core) BG solitons \cite
{review} and solitons in dual-core fibers without BG \cite{MakAsymm}.

In line with the above arguments, other basic characteristics of the new
solitons are expected to be on the same order of magnitude as those for the
recently observed BG solitons in the single-core fiber. In particular, the
soliton can be generated by a laser pulse of the duration $\sim 100$ ps,
having a fairly high peak power $\sim 5$ W (which is, however, still
sufficiently far from the optical-breakdown threshold in silica glass) and,
accordingly, the energy $\sim 500$ pJ . The soliton to be created will keep,
essentially, all this energy, self-compressing to the temporal width $%
\,_{\sim }^{<}\,\,50$ ps \cite{exper,exper2}.

Another crucial ingredient of the possible experiment is the necessary
length of the BG-equipped dual-core fiber. As it was mentioned above, for
the successful generation and detection of the gap soliton in the
single-core BG fiber, a $6$ cm fiber was sufficient. In fact, the
present-day techniques make it quite easy to fabricate a homogeneous
dual-core fiber of the length $\sim 1$ m, as well as to write a uniform BG
on it. Therefore, the experiment may be quite feasible in the fiber whose
length is on the order of $100$ characteristic soliton and coupling lengths
(both being $\sim 1$ cm, see above), which will be more than enough for the
most precise experiments.

Thus, experimental generation of the new solitons, to be theoretically
studied in the present work, is not going to be much harder than the recent
experiments reported in Refs. \cite{exper} and \cite{exper2}. The only
essentially new issue in the experiment may be a question if the input laser
pulse may be focussed, as usual, on the entrance face of one core only, or
it is necessary to split it, in a special fashion, between the two cores.
Although it may be premature here to discuss experimental technicalities in
such a detail, we notice that having the fiber length much longer than $z_{%
{\rm coupl}}\sim 1$ cm, see above, will provide enough room for the proper
redistribution of the power between the cores, so that the experiment will
scarcely be critically sensitive to details of launching the input pulse.

The rest of the paper is organized as follows. In section 2 we display {\em %
exact analytical} soliton solutions that can be found in the present model
with $c=0$, and results of simulations of their stability, which show that
they are stable in a broad parametric region. In the case $c\neq 0$, soliton
solutions can only be found numerically, which is done in section 3,
together with systematic simulations of their stability. It is found that,
depending on the value of frequency $\omega$ the solitons may remain stable
up to a large value $c=c_{{\rm max}}$. At $c>c_{{\rm max}}$, the soliton 
becomes unstable. This instability, however, does not destroy it, but, after
shedding  some radiation, it evolves into another member of the soliton
family.

In section 4 we directly simulate interactions between two solitons placed
initially at some distance from each other. It is found that the result of
the interaction strongly depends on the relative phase of the two solitons.
In particular, the interaction can easily generate moving solitons and leads
to spontaneous symmetry breaking.

\section{Exact soliton solutions and their stability}

Exact zero-velocity soliton solutions to Eqs. (\ref{u}) - (\ref{psi}) can
only be found in the particular case $c=0$. Starting with the usual ansatz, 
\begin{eqnarray}
u &=&U(x)\exp (-i\omega t),\,v=V(x)\exp (-i\omega t),  \label{uvansatz} \\
\phi &=&\Phi (x)\exp (-i\omega t),\,\psi =\Psi (x)\exp (-i\omega t)\,,
\label{phipsiansatz}
\end{eqnarray}
and following the pattern of the exact GMTM solutions \cite{AW}, we find 
\begin{eqnarray}
U(x) &\equiv &\left[ \frac{\left( \omega ^{2}-\lambda ^{2}-\mu ^{2}+\lambda
\kappa ^{2}\right) ^{2}+\mu ^{2}\kappa ^{4}}{\left( \omega ^{2}-\lambda
^{2}-\mu ^{2}\right) ^{2}}\right] ^{1/4}\,\,e^{+i\delta /2}\,A(x)\,,\, 
\nonumber \\
\,V(x) &\equiv &\left[ \frac{\left( \omega ^{2}-\lambda ^{2}-\mu
^{2}+\lambda \kappa ^{2}\right) ^{2}+\mu ^{2}\kappa ^{4}}{\left( \omega
^{2}-\lambda ^{2}-\mu ^{2}\right) ^{2}}\right] ^{1/4}\,\,e^{-i\delta
/2}\,B(x)\,,  \label{tilde}
\end{eqnarray}
where $\delta =$ $\tan ^{-1}\left( \kappa ^{2}\mu /\left( \omega
^{2}-\lambda ^{2}-\mu ^{2}+\lambda \kappa ^{2}\right) \right) $, and 
\begin{eqnarray}
A\,(x) &=&\sqrt{2/3}\left( \sin \theta \right) \,{\rm sech}\left( \eta
x\cdot \sin \theta -i\theta /2\right) \,,  \nonumber \\
B(x) &=&-\sqrt{2/3}\left( \sin \theta \right) \,{\rm sech}\left( \eta x\cdot
\sin \theta +i\theta /2\right) \,,  \label{UV}
\end{eqnarray}
\begin{eqnarray}
\Phi (x) &=&-\frac{\kappa \omega }{\omega ^{2}-\lambda ^{2}-\mu ^{2}}U+\frac{%
\kappa \left( \lambda +i\mu \right) }{\omega ^{2}-\lambda ^{2}-\mu ^{2}}V\,,
\nonumber \\
\Psi (x) &=&\frac{\kappa \left( \lambda -i\mu \right) }{\omega ^{2}-\lambda
^{2}-\mu ^{2}}U-\frac{\kappa \omega }{\omega ^{2}-\lambda ^{2}-\mu ^{2}}V\,.
\label{PhiPsi}
\end{eqnarray}
Here $\theta $, which takes values between $0$ and $\pi $, is an arbitrary
parameter of the soliton family. The frequency $\omega $ and inverse width $%
\eta $ of the soliton are determined, in terms of $\theta $, by equations 
\begin{equation}
\frac{\omega \left( \omega ^{2}-\lambda ^{2}-\mu ^{2}-\kappa ^{2}\right) }{%
\sqrt{\left( \omega ^{2}-\lambda ^{2}-\mu ^{2}+\lambda \kappa ^{2}\right)
^{2}+\mu ^{2}\kappa ^{4}}}\cdot {\rm sgn}\left( \omega ^{2}-\lambda ^{2}-\mu
^{2}\right) =\cos \theta \,,  \label{theta}
\end{equation}
\begin{equation}
\eta \equiv \left[ \frac{\left( \omega ^{2}-\lambda ^{2}-\mu ^{2}+\lambda
\kappa ^{2}\right) ^{2}+\mu ^{2}\kappa ^{4}}{\left( \omega ^{2}-\lambda
^{2}-\mu ^{2}\right) ^{2}}\right] ^{1/4}.  \label{x}
\end{equation}
It is relevant to note that these exact solutions resemble those found
earlier in a linearly coupled system of cubic and linear Ginzburg-Landau
(GL) equations \cite{we}; however, the exact solutions to the GL equations
exist as isolated ones, rather than in families, i.e., they do not contain
any arbitrary parameter.

Before proceeding to numerical search for solitons in the case $c\neq 0$, it
is necessary to address the stability of the exact analytical solutions
obtained above. The first nonrigorous stability analysis of GMTM solitons
was done using the variational approximation \cite{Tasgal}. It was predicted
that instability might occur when an internal parameter of GMTM solitons $%
\theta $, similar to that introduced above in Eqs. (\ref{UV}), exceeds a
certain critical value, which was close to $\pi /2$. Then, a rigorous
treatment of the stability problem for the GMTM system, based on the
consideration of its linearized version, was developed in Refs.~\cite{exact}%
. It was demonstrated there that the solitons with $\theta $ exceeding a
critical value, which is slightly larger than $\pi /2$, are unstable indeed.
However, the instability is weak, therefore it was hard to observe it in
direct simulations.

In this connection, it should be noted that while the results for the
solitons' stability in various models, obtained from the solution of the
corresponding eigenvalue problem for the linearized equations are more
rigorous (and usually are technically more difficult) than those produced by
direct simulations of the nonlinear equations, the latter results may be
more appropriate for the physical applications. Indeed, if the soliton is,
rigorously speaking, unstable but the instability is weak (as is the case
for GMTM), it may happen that neither direct simulations performed for a
limited evolution time (or propagation distance, depending on the particular
system) nor a real experiment in a finite-size sample will demonstrate the
instability, so that, in terms of real physics, the soliton should be
regarded as a {\it stable} object, in accord with the prediction of the
direct simulations, and despite the contradiction with the rigorous results. 
Solitons in the BG fiber may provide an example of 
this situation. In this case,  experimental 
results \cite{exper,exper2}, while being in good 
agreement with direct simulations, have not been able to demonstrate the 
sophisticated instability predicted on the basis of the linearized 
equations in Refs.\cite{exact}. On the other hand, it is necessary to 
mention that, although the physical value of the soliton's peak power 
in these experiments was quite high, the actually observed BG 
solitons may still be low-intensity ones from the viewpoint of the 
corresponding theoretical model. However, the above-mentioned 
``sophisticated instability'' takes place only for high-intensity 
solitons. Another  example that could be cited regarding the fact that
solitons' instability may sometimes be formal is provided by
``spinning'' (2+1)-dimensional solitons in media with the cubic-quintic
nonlinearity. As the analysis of the corresponding linearized problem shows,
the solitons with the ``spin'' $s=1$ are, strictly speaking, always
unstable against infinitesimal azimuthal perturbations which destroy the
cylindrical symmetry of the solitons. Nevertheless, if the size of the
``spinning'' soliton is large enough, the instability may be so weak that
the soliton may persist as a fairly robust object over several  diffraction 
lengths \cite{CQ}, thus having a fairly good chance to be observed 
in an experiment.

To test the stability of the exact soliton solutions given by Eqs. (\ref
{uvansatz}) through (\ref{x}), we simulated their evolution by means of the
split-step Fourier algorithm, imposing various asymmetric (sometimes
nonsmall) initial perturbations. A typical case is displayed in Fig.~1,
showing that after shedding off some radiation, the perturbed pulse readily
evolves into a member of the soliton family (in fact, the final soliton in
Fig. 1 acquires a very small velocity, because the asymmetric perturbation
has ``pushed'' it; moving solitons will be specially considered below). In
particular, an important finding is that, when $\lambda \neq 0$, Eq. (\ref
{theta}) gives rise to three distinct roots for $\omega $, of which only the
one with largest $|\omega |$ is found to produce a stable soliton. On the
other hand, there are two different roots for $\omega $ at $\lambda =0$, 
{\em both} leading to stable solitons.

We have also found that, similar to the GMTM solitons, a fundamental
property of the soliton family in our extended model is that a stable part
of the family is {\em limited}, $\theta \leq $ $\theta _{{\rm max}}$, where $%
\theta _{{\rm max}}$depends on $\kappa $ and $\lambda $. To analyze this in
detail, we set $\lambda =0$, focusing on the simplest and most fundamental
case when BG is present in the nonlinear core only, and the stability is
solely controlled by the coefficient of the linear coupling between the
nonlinear and linear cores. The stability border inside the soliton family, $%
\theta _{{\rm max}}(\kappa )$, was then sought for gradually increasing $%
\theta $ at a fixed value of $\kappa $. We started from $\theta =\pi /12$,
where the exact soliton is definitely stable, until we hit a value $\theta _{%
{\rm max}}$ that gave rise to instability. The instability, when it sets in,
causes straightforward decay of the soliton into radiation. We have thus
found that $\theta _{{\rm max}}($ $\kappa =0.01)=$ $\pi /1.7$, $\theta _{%
{\rm max}}($ $\kappa =1)=$ $\pi /2.0$, and $\theta _{{\rm max}}(\kappa
=100)= $ $\pi /1.8$, i.e., the dependence of the stability limit on $\kappa $
is fairly weak, $\theta _{{\rm max}}$ being close to that in the single-core
model, although the shapes of the exact solitons may be really different.

\section{Solitons in the model with $c\neq 0$}

The above consideration pertained to the limiting case $c=0$, when the exact
solutions were available. The next necessary step is to consider $c\neq 0$,
when no exact solution for the zero-velocity solitons could be found. We
therefore started by using the known relaxation algorithm \cite{recipes} in
order, first of all, to obtain stationary soliton solutions numerically from
the ordinary differential equations produced by the substitution of Eqs. (%
\ref{uvansatz}) and (\ref{phipsiansatz}) into Eqs. (\ref{u}) - (\ref{psi}).
By properly setting boundary conditions, it was always possible to obtain a
soliton solution for a given $\omega $.

A major objective here is to find out whether at fixed  values of all
parameters except $c$, there exists a maximum  value of $c$ above  which the
solitons are unstable. It has been found that, depending on the  value of $%
\omega$, there indeed exists $c_{{\rm max}}$ beyond which solitons  become
unstable. However the instability at $c>c_{{\rm max}}$ leads not to 
disappearance of solitons, but rather to their self-rearrangement  into a
slightly  different form.

A typical result is displayed in Fig. 2, with $\kappa =1,\lambda  =\mu  =0 $
and $\omega =1.6$. This value of $\omega $ was chosen since it  lies 
sufficiently deep inside the stability region at $c=0$. Our simulations 
show  that the value $c_{{\rm max}}$ is  very  large, $\approx 4.2$. As is
seen in Fig. 2b, at $c>c_{{\rm max}}$ the soliton becomes unstable and,
after  shedding some radiation, it evolves into another member of the 
soliton family.

A practically significant consequence of the above result is that values of $%
c$ close to $1$ (recall that $1$ group velocity in the nonlinear core)
definitely give rise to stable solitons. This inference is important for
experiments, because, in the most realistic case when both cores are made of
the same material, the group velocities in them are necessarily close.

In the case $\lambda \neq 0$ (when BG is written on the linear core too),
the relaxation algorithm also successfully generated stationary solitons.
Starting with these, we have found that the solitons are stable in a broad
parametric region, again including values of $c$ essentially exceeding $1$.
However, detailed analysis of the interplay of $\lambda $ with other
parameters is very cumbersome and is left aside.

\section{Interactions between solitons and generation of moving solitons}

Since the present model is nonintegrable, interactions between solitons may
be quite complex. The simplest approach to simulating these interactions is
to start from a superposition of two identical exact solitons placed
initially at a distance from each other with some phase difference $%
\Delta\varphi $. Results of the simulations, typical examples of which are
displayed in Fig. 3, are similar for different values of the soliton's
internal parameters and initial separation (provided that the solitons
overlap weakly), but they strongly depend on $\Delta \varphi $. In the case $%
\Delta \varphi =\pi $, the solitons, quite naturally, repel each other, cf.
the well-known fact that nonlinear Schr\"{o}dinger (NLS) solitons interact
repulsively when $\Delta \varphi =\pi $ \cite{KM}. Even if the initial
separation between the solitons is relatively large, the repulsion is strong
enough to lend the two initially quiescent solitons conspicuous velocities,
see Fig. 3a. In this case, eventual velocities are found to be $W_{\pm }=\pm
0.03$. Thus, these simulations not only shed light on the character of the
interaction between the solitons, but also provide a convenient way to
generate stable {\em moving }ones.

It is also interesting to compare the initial {\em energy} $E_{i}$ of each
soliton, defined as $\int_{-\infty }^{+\infty }\left( |u|^{2}+|v|^{2}+|\phi
|^{2}+|\psi |^{2}\right) \,dx$ (which is a dynamical invariant of the
model), and final values $E_{f}$ of the energy of the moving solitons. In
the case shown in Fig. 3a, $E_{f}/E_{i}=0.986$, i.e., about $1.5\%$ of the
initial energy is lost (into emission of radiation) as a result of the
interaction process. It should be stressed that moving solitons produced by
the interaction exhibit some internal vibrations, i.e., the solitons appear
with a weakly excited internal mode (the existence of internal modes in
stable GMTM solitons is a known fact \cite{Tasgal,exact}). It may also
happen that they capture some radiation which will be very slowly radiated
away in the course of very long evolution (which is not relevant for
experiments).

In the opposite case $\Delta \varphi =0$, the solitons attract each other,
which is similar to what is known for the NLS solitons. As is shown in Fig.
3b, they temporally merge into a single pulse, which later splits into two
moving solitons with small internal vibrations. In this case, a conspicuous
breaking of the initial symmetry between the two solitons is observed
(special care has been taken to check that it is not an artifact produced by
the numerical scheme). A plausible explanation is that the lump produced by
the strong temporary overlapping of the initially attracting solitons (see
Fig. 3b) is unstable against symmetry-breaking perturbations, the breaking
being incomplete since the solitons separate quickly enough. This conjecture
seems natural, as it is well known that various multisoliton states in the
NLS equation are strongly unstable in the case of attraction \cite{KM}, but,
of course, much more extensive simulations are necessary to check it in
detail.

This partial symmetry breaking can be characterized by the final/initial
energy ratios for the two solitons shown in Fig. 3b, which is found to be $%
E_{f}/E_{i}=0.892$ and $0.864$ for the left and right solitons,
respectively. In this case, a considerable share of the initial energy, $%
\approx 12\%$, is lost into radiation. The final velocities of the solitons
are $W_{\pm }=\pm 0.22$, i.e., there is no tangible symmetry breaking in
terms of the velocities. Note that $\left| W_{\pm }\right| $ are much larger
in this case than in the case $\Delta \varphi =\pi $.

In the case $\Delta \varphi =0$, there is another noteworthy aspect of the
symmetry breaking: the final solitons demonstrate an {\em internal}
asymmetry, characterized by the ratios of their partial energies, 
\[
\varepsilon _{- }= \int_{-\infty }^{+\infty }\left| u_{- }\right|
^{2}dx/\int_{-\infty }^{+\infty }\left| v_{- }\right| ^{2}dx,\,\,\,
\varepsilon _{+}= \int_{-\infty }^{+\infty }\left| v_{+ }\right|
^{2}dx/\int_{-\infty }^{+\infty }\left| u_{+ }\right| ^{2}dx, 
\]
where the positive (negative) subscript pertains to the right (left)
soliton. For the solitons shown in Fig. 3b, this ratio takes values $%
\varepsilon _{-}=0.470$ and $\varepsilon _{+}=0.465$. In accord with these
values, the final solitons, being intrinsically asymmetric, are, to a good
approximation, {\em mirror images} of each other.

We have also simulated the interaction of solitons with the initial phase
difference is $\Delta \varphi =\pi /2$. In this case the solitons repel each
other, about $0.7\%$ of the energy is lost into radiation, and the symmetry
breaking is much more conspicuous, with the final velocities being $%
W_{-}=-0.023$ and $W_{+}=0.019$. The stronger symmetry breaking in this case
can be easily understood, as the symmetry of the initial configuration,
which was taken as $u_{{\rm sol}}(x-\frac{1}{2}x_{0})+iu_{{\rm sol}}(x+\frac{%
1}{2}x_{0})$, $x_{0}$ being the initial separation between the solitons, is 
{\em not} compatible with the Eqs. (\ref{u})-(\ref{psi}) and is therefore
broken upon the propagation in a straightforward way.

\section{Conclusion}

In this paper, we have introduced a model consisting of two linearly coupled
cores, one having the Kerr nonlinearity and the other being linear. A Bragg
grating is written on the nonlinear core, while the linear one may or may
not be equipped with the grating. The model allows to considerably extend
the previously studied family of the Bragg-grating solitons. Exact solutions
were found for zero-velocity solitons in a limiting case when the
group-velocity terms are absent in the equations for the linear core, while
in the general case solitons were found numerically. The main issue is their
stability. We have found a nontrivial stability limit for them in terms of 
an internal parameter of the soliton family. Depending on the frequency $%
\omega$, the solitons may remain stable up to quite large values of the
group velocity in  the linear core. This strongly suggests that stable
solitons can indeed be generated experimentally in dual-core systems, with
the cores made of the same material. The vast stability region for the
zero-velocity solitons in the dual-core model, found in this work, suggests
a possibility to experimentally look for the corresponding localized states
with the fully trapped light. Interactions of initially separated solitons
were investigated too, showing a considerable spontaneous symmetry breaking
in the case when the solitons attract each other, which may be a result of a
natural instability against symmetry-breaking perturbations. The interaction
always results in appearance of stable moving solitons.

\smallskip J.A.~thanks J.M. Soto-Crespo for useful discussions.
B.A.M.~acknowledges hospitality of the School of Electrical Engineering and
Communications at the University of New South Wales (Sydney), and of the
School of Physics at the University of Sydney.

\newpage

\section*{Figure Captions}

Fig.~1. Evolution of an asymmetrically perturbed soliton when $c=0$ (only
the $u$ -component is shown). The other parameters are $\lambda =\mu =0$, $%
\kappa =1$ , and the soliton's internal parameter (see Eqs. \ref{UV}) is $%
\theta =\pi /3$.

Fig.~2. Evolution of solitons at values of $c$ slightly below and above $c_{%
{\rm max}}$: (a) $c=4.1$; (b) $c=4.3$. The other parameters are $\kappa =1$, 
$\lambda =\mu =0$, and $\omega =1.6$.

Fig.~3. Interaction of two identical solitons with $\theta =\pi /3$, placed
initially at a distance $8$ with two different values of the initial phase
difference between the solitons: (a) $\Delta \varphi =\pi $ (repulsion); (b) 
$\Delta \varphi =0$ (attraction). The other parameters are $\kappa =1$, $%
\lambda =\mu =0$, and the evolution time is $400$.

\newpage 
\begin{figure}[th]
\centerline{\epsfysize=15cm\epsffile{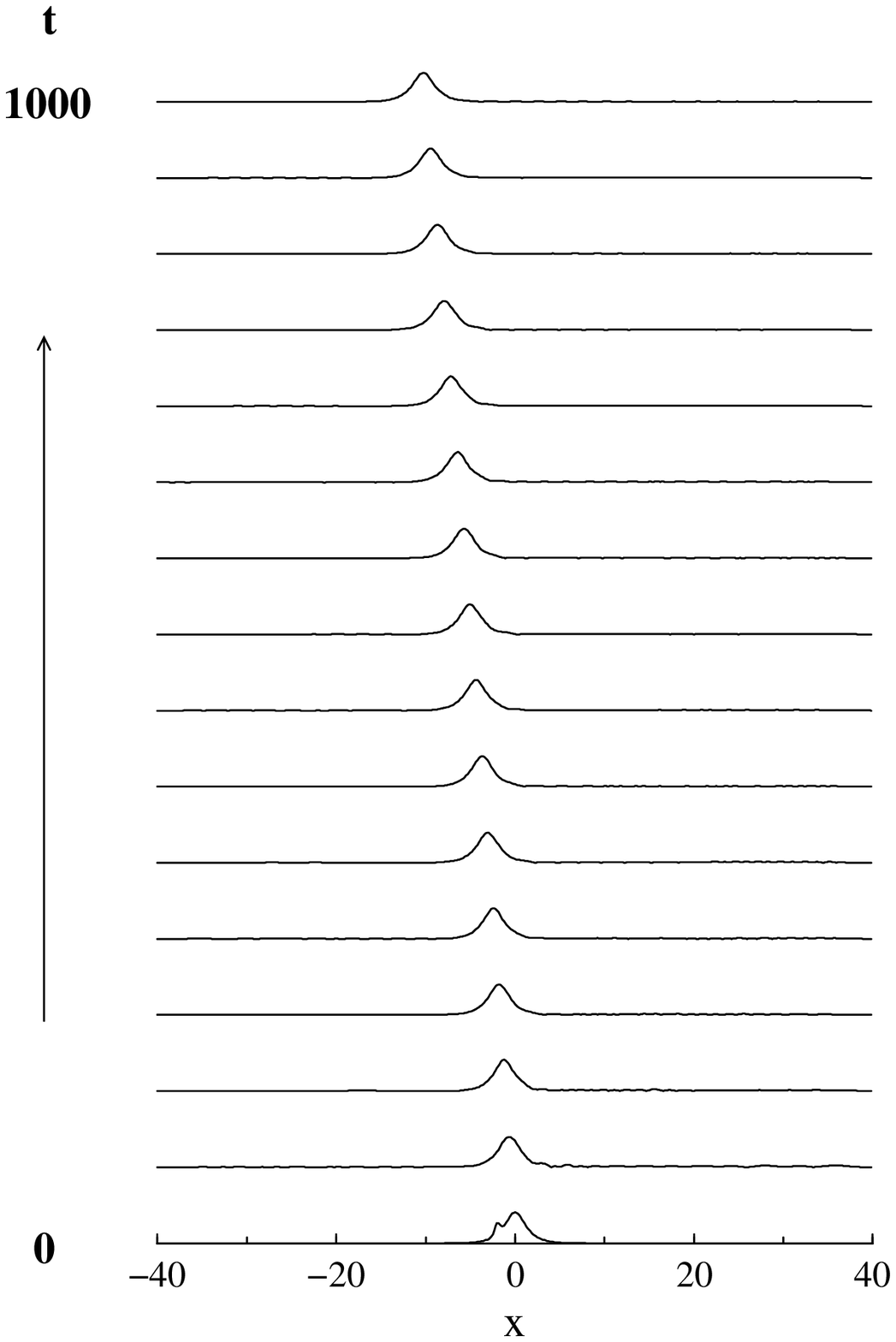}} 
Fig.~1.~Javid Atai and Boris A.~Malomed
\end{figure}

\newpage 
\begin{figure}[th]
\centerline{\epsfysize=15cm\epsffile{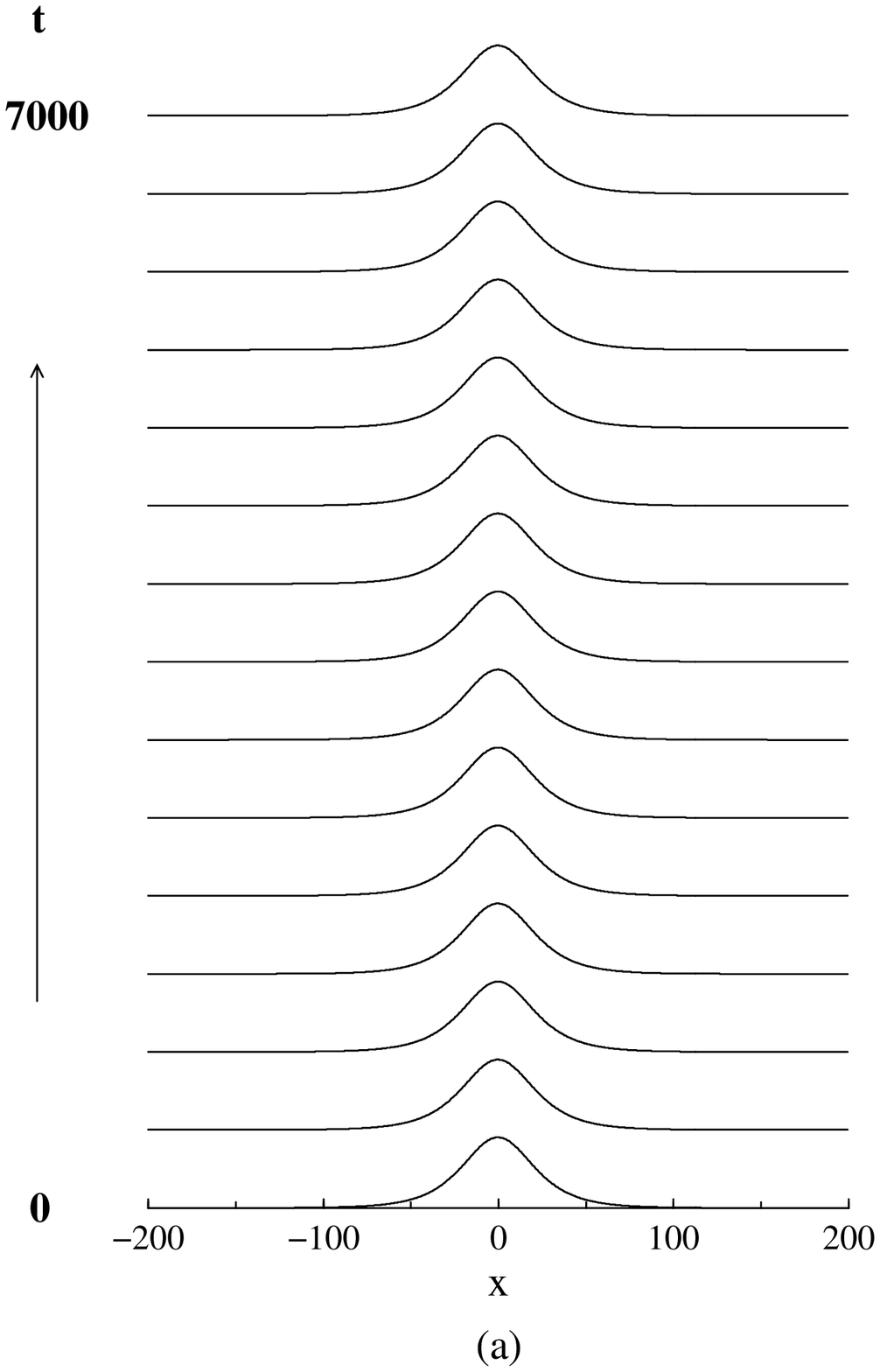}} 
Fig.~2.~Javid Atai and Boris A.~Malomed
\end{figure}
\newpage 
\begin{figure}[th]
\centerline{\epsfysize=15cm\epsffile{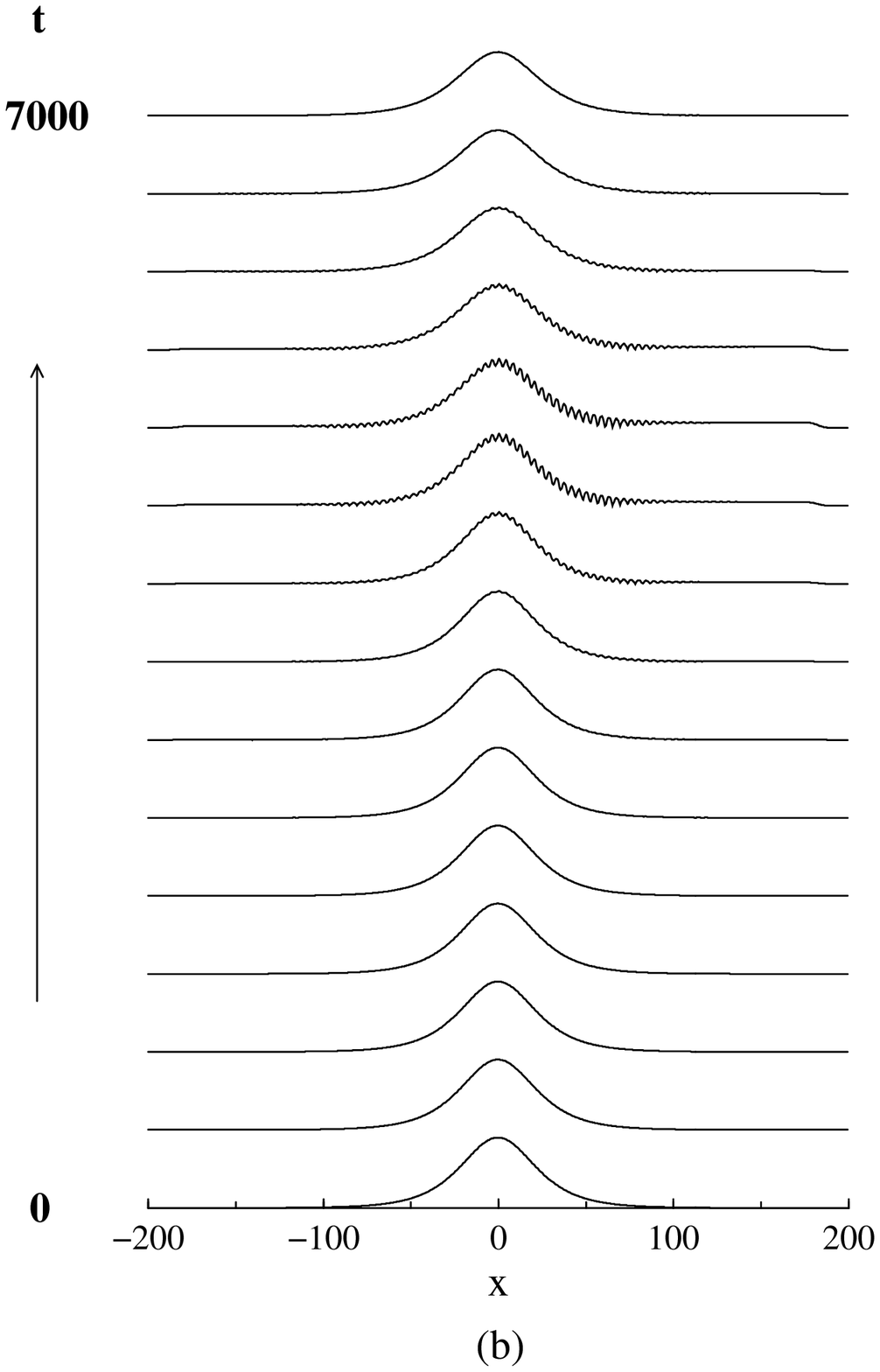}} 
Fig.~2.~Javid Atai and Boris A.~Malomed
\end{figure}

\newpage 
\begin{figure}[th]
\centerline{\epsfxsize=269.47307pt\epsfysize=15cm\epsffile{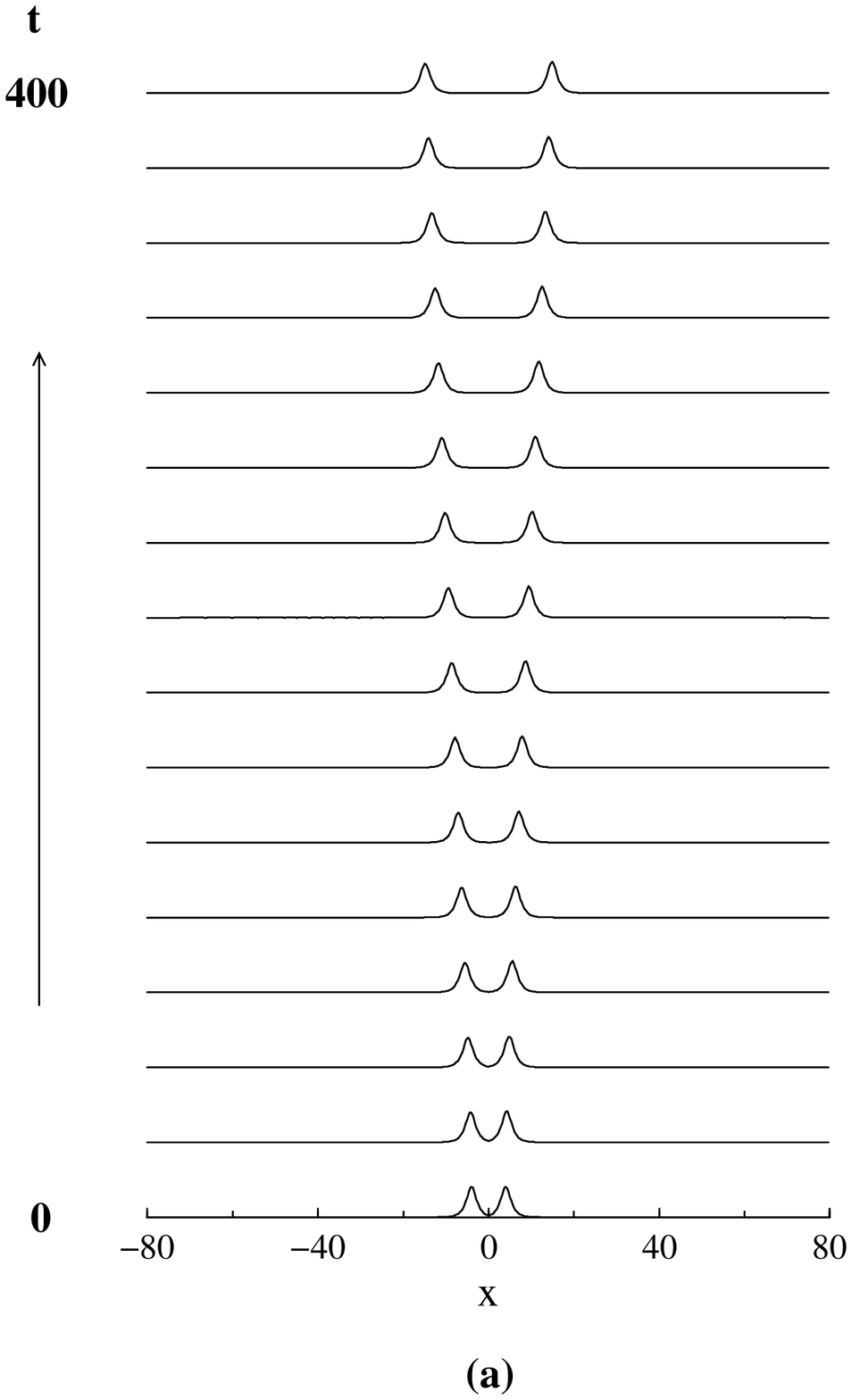}} 
Fig.~3.~Javid Atai and Boris A.~Malomed
\end{figure}
\newpage 
\begin{figure}[th]
\centerline{\epsfxsize=269.47307pt\epsfysize=15cm\epsffile{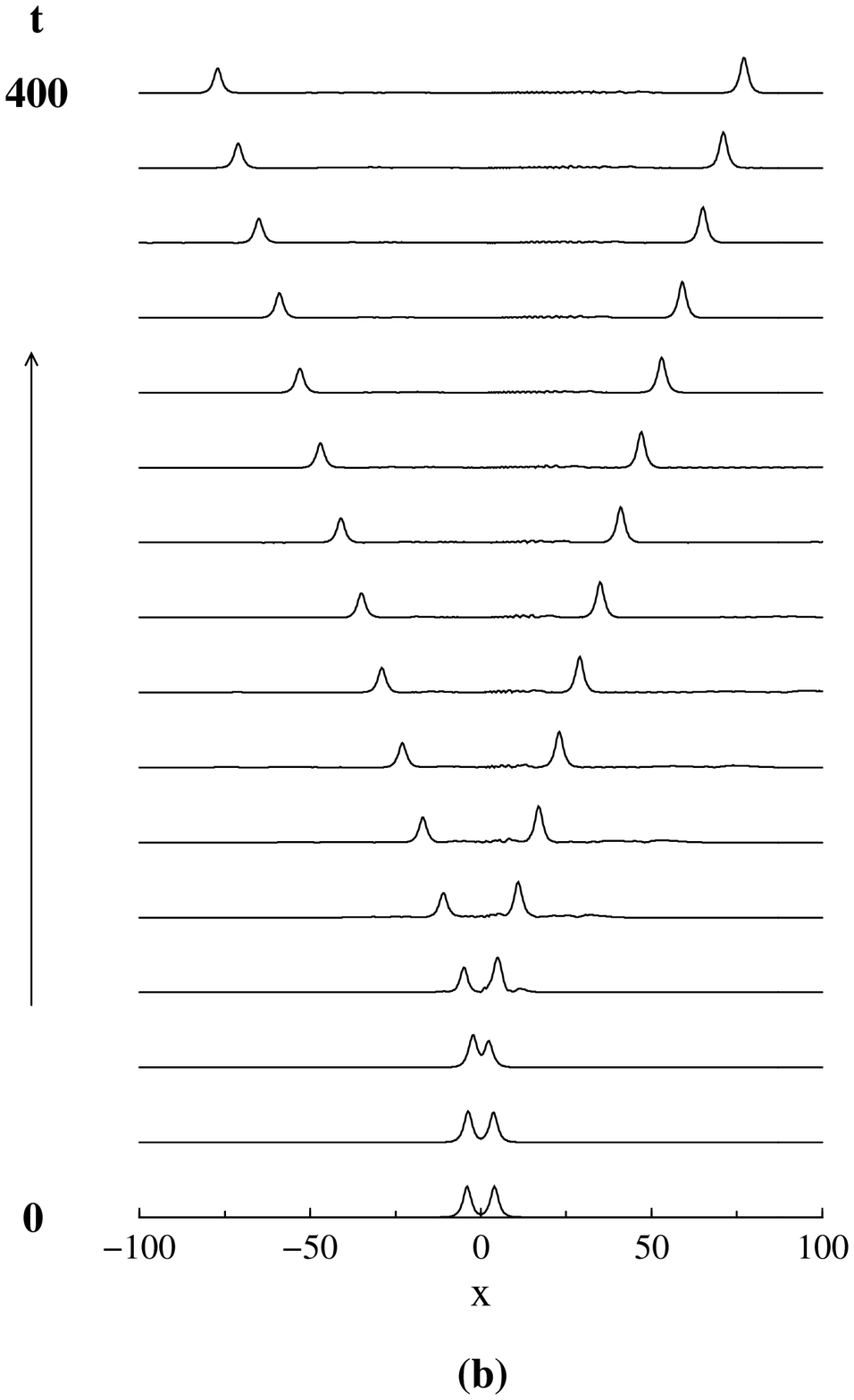}} 
Fig.~3.~Javid Atai and Boris A.~Malomed
\end{figure}

\end{document}